\documentclass[a4paper]{article}

\usepackage[dvips]{epsfig}
\usepackage{amsmath}
\usepackage{amstext}
\usepackage{amssymb}

\def\thebibliography#1{\section*{\normalsize \bf References 
 }\list
 {[\arabic{enumi}]}{\settowidth\labelwidth{[#1]}\leftmargin\labelwidth
 \advance\leftmargin\labelsep
 \usecounter{enumi}}
 \def\newblock{\hskip .11em plus .33em minus .07em}
 \sloppy\clubpenalty4000\widowpenalty4000
 \sfcode`\.=1000\relax}

\begin{document}
\sloppy
\twocolumn[
\fbox{\small PREPRINT: \hspace{1cm} Date: 05/25/1998 \hspace{1cm} Rev: 1.1
  \hspace{1cm} Status: to be published in Phys. Stat. Sol. B 208 \hspace{1cm}}
\begin{center}   
\large \bf
  Ferromagnetism within the periodic Anderson model:\\
  A new approximation scheme
\end{center}
\vspace{-3mm}
\begin{center} 
  D. Meyer\addtocounter{footnote}{1}\footnotemark and  W. Nolting
\end{center}
\vspace{-6mm}

\begin{center} \small \it 
  Institut f\"ur Physik, 
  Humboldt-Universit\"at zu Berlin,
  10115 Berlin,
  Germany
\end{center}

\begin{center} 
  G. G. Reddy and A. Ramakanth
\end{center}
\vspace{-6mm}

\begin{center} \small \it 
  Department of Physics,
  Kakatiya-University,
  Warangal-506009,
  India
\end{center}
\vspace{2mm}

\begin{center}
\parbox{141mm}{ We introduce a new approach to the periodic Anderson
model (PAM) that allows a detailed investigation of the magnetic
properties in the Kondo as well as the intermediate valence regime. Our
method is based on an exact mapping of the PAM onto an effective medium
strong-coupling Hubbard model. For the latter, the so-called spectral
density approach (SDA) is rather well motivated since it is based on
exact results in the strong coupling limit.  Besides the $T=0$ phase
diagram, magnetization curves and Curie temperatures are presented and
discussed with help of temperature-dependent quasiparticle densities of
state. In the intermediate valence regime, the hybridization gap plays a
major role in determining the magnetic behaviour. Furthermore, our
results indicate that ferromagnetism in this parameter regime is not
induced by an effective spin-spin interaction between the localized
levels mediated by conduction electrons as it is the case in the Kondo
regime.  The magnetic ordering is rather a single band effect within an
effective $f$-band.  \vspace{4mm}

}
\end{center}
\vspace{8mm} 
]
\footnotetext[2]{corresponding author: Dietrich.Meyer@physik.hu-berlin.de}
\section{Introduction}
The lanthanides and actinides, and their compounds, show a great variety
of interesting physical properties. Some of these have to be ascribed to
the most unexpected and least understood phenomena in condensed matter
physics. Probably, the most diverse physical characteristics are found
in the heavy-fermion (HF) and intermediate valence (IV) materials. A
comprehensive review is given in ref. \cite{GS91}.

Prototypes are the Cerium (Ce) and Uranium (U) intermetallics. In these
materials, incompletely filled inner $4f$-shells (in cerium, $5f$ in
uranium) are responsible for the unusual physical properties. Worth to
mention are, of course, the name-giving heavy fermion phenomenum which
is characterized by an enormous effective mass of crystal electrons and
the anomalous superconductivity which is called anomalous due to its
coexistence with magnetic ordering. The magnetic phase diagrams of these
materials are quite extraordinary in their large variety of different
phases, ranging from simple paramagnetic states to various kinds of
magnetic ordering.

Ferromagnetism is found in several Kondo lattice and HF materials. For
instance, in $CeSi_x$ phase transitions as functions of the $Si$
concentration were observed \cite{HWA92}, or in the heavy fermion
compound $CePd_2Ga_3$, ferromagnetic ordering breaks down as function of
external pressure \cite{BHE97}.  For $CeSi_x$, the magnetic phase
diagram with respect to the $Si$ concentration $x$ seems quite clear,
but some effects accompaying the magnetic ordering, like the irregular
resistivity behaviour, are not yet understood.  For the second material,
$CePd_2Ga_3$, not even the magnetic phase diagram is settled yet. The
pressure-induced suppression of ferromagnetism could be followed by some
other magnetically ordered phase.  Besides of the above mentioned, a lot
more materials showing ferromagnetic ordering were found.

Most attempts to understand HF- or IV-materials theoretically are based
on the Anderson model \cite{And61}. This model describes a system of
uncorrelated conduction electrons which hybridizes with either a single
localized electronic level (single impurity Anderson model, SIAM) or a
lattice of localized levels (periodic Anderson model, PAM).

Many approximation schemes for these models have been proposed. At least
for the SIAM, there is enough confidence in some of these approximations
that most of the results are widely accepted. The so-called non-crossing
approximation (NCA)\cite{PG89} has proven to be in many aspects
equivalent to quasi-exact QMC calculations \cite{PCJ93}. In addition, by
means of Bethe ansatz \cite{TW83} and renormalization group theories,
many exact results could be obtained.

For the PAM, the situation is not yet comparable. Even though many
approximation schemes exist, beginning with the mean-field-approximation
\cite{LM78}, second order perturbation theory in $U$ \cite{HC96} or in
the hybridization \cite{VGR92,VGR94}, a modification of the non crossing
approximation (LNCA) \cite{GPK88,Gre88}, and various dynamic mean-field
theories \cite{GKKR96, TJF97}, none of these are completely satisfactory
in all aspects of the PAM. Some of the approximations, namely the LNCA
\cite{GPK88,Gre88} are restricted to the Kondo regime of the Anderson
model. In this parameter regime, denoted by a half-filled low-lying
localized state, it is possible to map the PAM onto the $s$-$f$ model in
which the localized states are reduced to their spin degrees of freedom
\cite{SW66}.

So, despite of all efforts done in this field, no fully satisfactory
theoretical understanding of the $4f$-materials is developed yet. Even
on the experimental side, not everything has been clarifed. For
instance, it is currently intensely discussed, whether the Kondo
resonance should be seen in photoemission studies on periodic crystals
\cite{MGB96,Huef96,Aea97}. But these disputs will probably be settled in
the near future.\\ For the theoretical side, it still seems neccessary
to develope better theories which help to understand at least partial
aspects of the PAM.

In this paper, we will introduce a novel approximation scheme for the
periodic Anderson model using an exact mapping of the PAM onto an
effective Hubbard model. The model parameters within this effective
model will be such that results obtained in a strong-coupling
perturbational theory are significant. These motivate the spectral
density approach (SDA) \cite{Nol72,NB89,HN97a} to solve the effective
Hubbard model. This approximation scheme has proven, at least for the
Hubbard model, to be trustworthy concerning the magnetic properties of
the system.\\ This approximation scheme will allow to solve the PAM for
a wide range of system parameters, the crossover from the Kondo limit to
the intermediate valence region will be subject of our
investigation. Although due to numerical simplicity, all our results
were obtained under the assumption of a $\vec{k}$-independent
selfenergy, this method is in principle not restricted to the local
approximation as dynamical mean-field approaches are. Furthermore, even
if based on exact results obtained in the limit $U\rightarrow\infty$, it
is not of perturbative character, therefor not restricted to this limit.
The major drawback is the neglection of quasiparticle damping, and
therewith the impossibility to show fermi liquid behaviour.

We will develope our approximation scheme in the next section of the
paper. In section three, the results obtained with this method
concerning ferromagnetism in the periodic Anderson model are presented.

\section{Theory}
\subsection{Model Hamiltonian and its many-body problem}
Starting point is the periodic Anderson Hamiltonian
\begin{align}
  \label{hamiltonian}
    H =&\sum_{i,j,\sigma} T_{i,j}s_{i,\sigma}^{\dagger}s_{j,\sigma} +
    \sum_{i,\sigma} \epsilon_f f_{i,\sigma}^{\dagger}f_{i,\sigma} +\\ &V
    \sum_{i,\sigma} (f_{i,\sigma}^{\dagger}s_{i,\sigma} +
    s_{i,\sigma}^{\dagger}f_{i,\sigma} ) + \frac{1}{2} U \sum_{i,\sigma}
    n_{i,\sigma}^{(f)}n_{i,-\sigma}^{(f)}\nonumber
\end{align}
$s_{i,\sigma}$ ($f_{i,\sigma}$) and $s_{i,\sigma}^{\dagger}$
($f_{i,\sigma}^{\dagger}$) are the annihilation and creation operators
for an electron in a non-degenerate conduction band state (localized
$f$-state), and $n_{i,\sigma}^{(f)}=f_{i,\sigma}^{\dagger}f_{i,\sigma}$
is the occupation number operator for the $f$-states. The hopping
integral
$T_{i,j}=\frac{1}{N}\Sigma_{\vec{k}}e^{-i\vec{k}(\vec{R}_i-\vec{R}_j)}
\epsilon(\vec{k})$ describes the propagation of free, i.e. unhybridized
conduction electrons from site $j$ to site $i$, $\epsilon_f$ is the
position of the free $f$-level relative to the center of mass of the
conduction band density of states. The hybridization $V$ is taken as a
real constant, and finally $U$ is the Coulomb repulsion between two
$f$-electrons on the same lattice site.  By use of the commutators
\begin{align}
  [s_{i,\sigma},H]_- &=\sum_m T_{i,m} s_{m,\sigma} +V f_{i,\sigma}\\
  [f_{i,\sigma},H]_- &= \epsilon_f f_{i,\sigma} +V s_{i,\sigma}+U
  f_{i,\sigma}n_{i,-\sigma}^{(f)}
\end{align}
we find for the single-$s$,$f$ electron Green's functions
\begin{align}
  \label{fgf}
  \langle\!\langle f_{i,\sigma};f_{j,\sigma}^{\dagger}\rangle\!\rangle_E
  &= \frac{1}{N}\sum_{\vec{k}}e^{-i \vec{k} ( \vec{R}_i -\vec{R}_j)}
  \langle\!\langle
  f_{\vec{k},\sigma};f_{\vec{k},\sigma}^{\dagger}\rangle\!\rangle_E \\
  &= \frac{1}{N}\Sigma_{\vec{k}} e^{-i \vec{k} ( \vec{R}_i -\vec{R}_j)}
  G_{\vec{k},\sigma}^{(f)}(E)\nonumber\\
  \label{sgf}
  \langle\!\langle s_{i,\sigma};s_{j,\sigma}^{\dagger}\rangle\!\rangle_E
  &= \frac{1}{N} \sum_{\vec{k}}e^{-i \vec{k} ( \vec{R}_i -\vec{R}_j)}
  \langle\!\langle
  s_{\vec{k},\sigma};s_{\vec{k},\sigma}^{\dagger}\rangle\!\rangle_E\\ &=
  \frac{1}{N}\Sigma_{\vec{k}} e^{-i \vec{k} ( \vec{R}_i -\vec{R}_j)}
  G_{\vec{k},\sigma}^{(s)}(E)\nonumber
\end{align}
the following equations of motion:
\begin{eqnarray}
    \label{eqmo1a}
\lefteqn{ \sum_{m}(E\delta_{i,m}-T_{i,m}) \langle\!\langle
s_{m,\sigma};s_{j,\sigma}^{\dagger}\rangle\!\rangle_E =}\\ &&
\hspace{18ex} \hbar \delta_{i,j}+ V \langle\!\langle
f_{i,\sigma};s_{j,\sigma}^{\dagger}\rangle\!\rangle_E \nonumber
   \end{eqnarray}
\begin{eqnarray}   
\label{eqmo1b}
    \lefteqn{(E-\epsilon_f) \langle\!\langle
    f_{i,\sigma};f_{j,\sigma}^{\dagger}\rangle\!\rangle_E =} \\ &&\hbar
    \delta_{i,j} + V \langle\!\langle
    s_{i,\sigma};f_{j,\sigma}^{\dagger}\rangle\!\rangle_E + U
    \langle\!\langle
    f_{i,\sigma}n_{i,-\sigma}^{(f)};f_{j,\sigma}^{\dagger}\rangle\!\rangle_E\nonumber
\end{eqnarray}
After Fourier transformation these equations read:
\begin{eqnarray}
  \label{eqmo2a}
  \lefteqn{(E-\epsilon(\vec{k})) G_{\vec{k},\sigma}^{(s)}(E) = \hbar + V
  \langle\!\langle
  f_{\vec{k},\sigma};s_{\vec{k},\sigma}^{\dagger}\rangle\!\rangle_E}\\
  \label{eqmo2b}
  \lefteqn{(E-\epsilon_f) G_{\vec{k},\sigma}^{(f)}(E) =}\\ &&
  \hspace{10ex}\hbar + V \langle\!\langle
  s_{\vec{k},\sigma};f_{\vec{k},\sigma}^{\dagger}\rangle\!\rangle_E + U
  D_{\vec{k},\sigma}(E)\nonumber
\end{eqnarray}
The ``higher'' Green's function $D_{\vec{k},\sigma}(E)$,
\begin{equation}
  \label{highd}
  D_{\vec{k},\sigma}(E)=\frac{1}{N}
  \sum_{\vec{p},\vec{q}}\langle\!\langle f_{\vec{p},-\sigma}^{\dagger}
  f_{\vec{q},-\sigma}f_{\vec{p}+\vec{k}-\vec{q},\sigma};f_{\vec{k},\sigma}^{\dagger}
  \rangle\!\rangle_E
\end{equation}
prevents a direct solution of the equations of motion. However, the
``mixed'' function in (\ref{eqmo2a}) and (\ref{eqmo2b}), respectively,
can easily be eliminated using the respective equation of motion:
\begin{equation}
  \begin{split}
    \langle\!\langle
    s_{\vec{k},\sigma};f_{\vec{k},\sigma}^{\dagger}\rangle\!\rangle_E &=
    \frac{V}{E-\epsilon(\vec{k})} G_{\vec{k},\sigma}^{(f)}(E)\\ &\equiv
    \langle\!\langle
    f_{\vec{k},\sigma};s_{\vec{k},\sigma}^{\dagger}\rangle\!\rangle_E
  \end{split}
\end{equation}
So we are left with the following set of equations of motion:
\begin{gather}
  \begin{split}
    \label{eqmo3a}
    (E-\epsilon(\vec{k}))& G_{\vec{k},\sigma}^{(s)}(E) =\\ & \hbar +
    \frac{V}{E-\epsilon(\vec{k})} G_{\vec{k},\sigma}^{(f)}(E)
  \end{split}\\
  \begin{split}
    \label{eqmo3b}
    (E-\epsilon_f-\frac{V^2}{E-\epsilon(\vec{k})} )&
    G_{\vec{k},\sigma}^{(f)}(E) =\\ & \hbar+ U D_{\vec{k},\sigma}(E)
  \end{split}
\end{gather}
which tells us that the determination of the $f$-Green's function
(\ref{fgf}) solves the problem.

The introduction of a selfenergy $\Sigma_{\vec{k},\sigma}(E)$ by
\begin{equation}
  \label{selfenergy}
  \Sigma_{\vec{k},\sigma}(E) G_{\vec{k},\sigma}^{(f)}(E)= U
  D_{\vec{k},\sigma}(E)
\end{equation}
allows a formal solution of the equations of motion (\ref{eqmo3a}) and
(\ref{eqmo3b}):
\begin{align}
  \label{gf_formal}
    G_{\vec{k},\sigma}^{(s)}(E) &= \hbar
    \frac{E-\Sigma_{\vec{k},\sigma}(E)-\epsilon_f}
    {(E-\Sigma_{\vec{k},\sigma}(E)-
    \epsilon_f)(E-\epsilon(\vec{k}))-V^2}\\ G_{\vec{k},\sigma}^{(f)}(E)
    &=
    \frac{\hbar}{E-\epsilon_f-\frac{V^2}{E-\epsilon(\vec{k})}-\Sigma_{\vec{k},\sigma}(E)}
\end{align}
From these we derive the quasiparticle densities of states (QDOS):
\begin{align}
  \label{rho.f}
  \rho_{s,\sigma}(E)&=\frac{1}{\hbar N} \sum_{\vec{k}}
  \left(-\frac{1}{\pi} \Im G_{\vec{k},\sigma}^{(s)}(E+i0^+)\right) \\
  \rho_{f,\sigma}(E)&=\frac{1}{\hbar N}\sum_{\vec{k}}
  \left(-\frac{1}{\pi} \Im G_{\vec{k},\sigma}^{(f)}(E+i0^+)\right)
\end{align}
In the case of a local selfenergy,
\begin{equation}
  \label{M_local}
  \Sigma_{\vec{k},\sigma}(E)\equiv \Sigma_{\sigma}(E)
\end{equation}
simple manipulations yield the following structures of the densities of
states:
\begin{align}
    \rho_{s,\sigma}(E)=&\rho_0\left(E-\frac{V^2}{E-\Sigma_{\sigma}(E)-\epsilon_f}\right)\\
  \begin{split}
    \rho_{f,\sigma}(E)=&
    \frac{V^2}{(E-\Sigma_{\sigma}(E)-\epsilon_f)^2}\\ &\times
    \rho_0\left(E- \frac{V^2}{E-\Sigma_{\sigma}(E)-\epsilon_f}\right)
  \end{split}
\end{align}
with $\rho_0(E)=\frac{1}{N} \Sigma_{\vec{k}}\hbar
\delta\left(E-\epsilon(\vec{k})\right)$ being the free $s$-electron
Bloch density of states.

The full problem is then obviously solved as soon as we have found the
selfenergy $\Sigma_{\vec{k},\sigma}(E)$, defined by
(\ref{selfenergy}). Our approximation scheme is developed in the next
two sections.
\subsection{Effective medium approach}
The basic idea of the approximation is a mapping of the periodic
Anderson model onto an effective medium Hubbard model. This effective
Hamiltonian will have an energy-dependent one-particle term, containing
the influence of the hybridization of the localized states with the
conduction band and a Hubbard-like Coulomb term for the respective
quasiparticles and has the following structure:
\begin{equation}
  \label{effham}
  \tilde{H}(\eta)=\sum_{\vec{k},\sigma} \Delta_{\vec{k}}(\eta)
  a_{\vec{k},\sigma}^{\dagger}a_{\vec{k},\sigma} + \frac{1}{2} U
  \sum_{i,\sigma} n_{i,\sigma}^{(a)}n_{i,-\sigma}^{(a)}
\end{equation}
with
\begin{align}
  \label{effmed}
  \Delta_{\vec{k}}(\eta) &= \frac{V^2}{\eta-\epsilon(\vec{k})}+\epsilon_f\\
  \label{effmed1}
  \Delta_{i,j}(\eta)&=\frac{1}{N} \sum_{\vec{k}} \Delta_{\vec{k}}(\eta)
  e^{i\vec{k}(\vec{R}_i-\vec{R}_j)}
\end{align}
being the effective band dispersion and
$n_{i,\sigma}^{(a)}=a_{i,\sigma}^{\dagger}a_{i,\sigma}$. $\eta$ has to
be understood as a parameter with the dimension of energy. Throughout
the paper, we will denote quantities which are defined in the effective
medium, with a tilde (\~{}). All these quantities depend on the
parameter $\eta$ which will often be omitted for readability.  The
Green's function for the system described by (\ref{effham}),
$\tilde{G}_{\vec{k},\sigma}^{(\eta)}(E)=\langle\!\langle
a_{\vec{k},\sigma};a_{\vec{k},\sigma}^{\dagger}\rangle\!\rangle_E$ can
be determined by the equation of motion:
\begin{equation}
  \label{eomeffmed}
    \left(E-\Delta_{\vec{k}}(\eta)  \right)
    \tilde{G}_{\vec{k},\sigma}^{(\eta)}(E) =  \hbar + U \tilde{D}_{\vec{k},\sigma}^{(\eta)}(E)
\end{equation}
The higher Green's function $\tilde{D}_{\vec{k},\sigma}^{(\eta)}(E)$ is defined analogously to
(\ref{highd}) and can be expressed in terms of a properly defined selfenergy:
\begin{equation}
  \tilde{\Sigma}_{\vec{k},\sigma}^{(\eta)}(E) \tilde{G}_{\vec{k},\sigma}^{(\eta)}(E)= U \tilde{D}_{\vec{k},\sigma}^{(\eta)}(E)
\end{equation}
By comparision of the two equations of motion (\ref{eqmo3b}) and
(\ref{eomeffmed}), it follows that 
\begin{equation}
  \label{equivgreen}
  \left. \tilde{G}_{\vec{k},\sigma}^{(\eta)}(E)\right|_{\eta=E} = G_{\vec{k},\sigma}^{(f)}(E)
\end{equation}
and equivalently for the respective selfenergies
\begin{equation}
  \label{equivself}
  \left.\tilde{\Sigma}_{\vec{k},\sigma}^{(\eta)}(E)\right|_{\eta=E}=\Sigma_{\vec{k},\sigma}(E)
\end{equation}
Thus, by solving the effective problem defined via (\ref{effham}) for
all values of $\eta$, one can obtain the $f$-Green's function of the
periodic Anderson model. The original problem has been mapped onto an
``effective'' Hubbard problem.

Since the Hubbard model, at least in more than one dimension, is not yet
solved exactly, the advantage of the mapping is not seen
immediately. But a closer analysis of the effective medium will show
that the mapping allows for a rather well motivated approximation.

\begin{figure}[htb]
  \begin{center}
    \epsfig{file=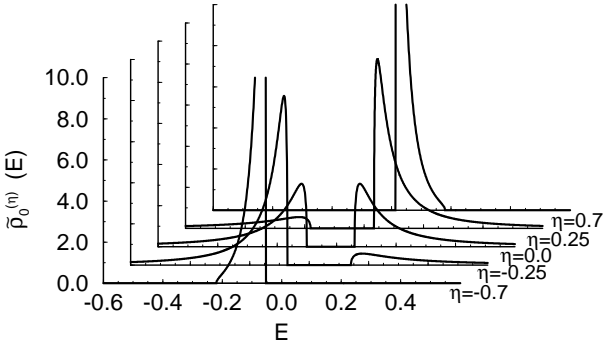, width=8cm}
    \caption{Effective density of states for different values of
      $\eta$. The free Bloch density of states is taken to be semielliptic
      of unit width and center of gravity at $E=0$. The hybridization is $V=0.2$ and the $f$-level
      $e_f=-0.02$} 
    \label{fig:bdos}
  \end{center}
\end{figure}

The ``free'' band dispersion $\Delta_{\vec{k}}(\eta)$ changes with
$\eta$ being, however, for reasonable $V$-values always very narrow. The
respective ``free'' density of states
\begin{equation}
  \label{effbdos}
  \begin{split}
  \tilde{\rho}_{0}^{(\eta)}(E)&=\frac{1}{N} \sum_{\vec{k}} \hbar \delta(E-\Delta_{\vec{k}}(\eta))\\
  &=\frac{V^2}{(E-\epsilon_f)^2}\rho_0\left(\eta-\frac{V^2}{E-\epsilon_f}\right)
  \end{split}
\end{equation}
is plotted in figure \ref{fig:bdos} for some typical examples of $\eta$. 
The effective dispersion $\Delta_{\vec{k}}(\eta)$ diverges for
$\epsilon(\vec{k})=\eta$, i.e. when $\eta$ falls into the band region
($\min \epsilon (\vec{k}) \leq \eta \leq \max \epsilon(\vec{k})$). The
resulting density of states extends from $-\infty$ to
$\infty$. But since $\tilde{\rho}_0^{(\eta)}(E) \sim \frac{1}{E^2}$, the
respective weight vanishes. Thus, the effective width $\tilde{W}$ is very small
scaling with the hybridization $V$ which must be considered for all
realistic situations as a small parameter, at least very much smaller
than the Coulomb interaction $U$. The ``Hubbard problem'' which is
given by the effective Hamiltonian (\ref{effham}) is therefore to be
ascribed to the strong coupling regime: $\frac{U}{\tilde{W}}\ggg 1$.

At this point, the advantage of the mapping onto the
effective model is clear. We have replaced the periodic Anderson model,
which is, though a minimal, still a two-band model with an effective
single band Hubbard model, for which some standard approximation methods 
exist. By construction of the effective medium, the single band model
turns out to belong to the strong coupling regime, thus results obtained 
in a $\frac{W}{U}$ perturbational theory \cite{HL67} can help
constructing an appropriate approximation scheme
\subsection{Spectral density approach}
We use a selfconsistent spectral density approach
(SDA)\cite{Nol72,NB89,HN97a} to find an approximate solution of the
``effective'' Hubbard problem defined by the Hamiltonian
$\tilde{H}(\eta)$ in (\ref{effham}). The method is based on a physically
motivated ansatz for the single-electron spectral density. Its main
advantages are the physically simple concept and the non-perturbative
character being not restricted to Fermi-systems but also working for
Bose- and even classical systems \cite{NO83,Cea84,Bea86}. Recent
applications of the SDA concern the attractive ($U<0$) Hubbard model
\cite{SPR96}, the $t-J$ model \cite{Mas93}, the magnetism and electronic
structure of systems of reduced dimensions as thin films and surfaces
\cite{PN96,PN97a,PN97b}. It is also used for the investigation of
high-$T_c$ superconductivity \cite{BE95,MEHM95}. In this paper we apply
the SDA to the ``effective medium''-Hubbard model (\ref{effham}).

The single-electron spectral density is defined by
\begin{multline}
  \label{spec_dens}
  \tilde{S}_{\vec{k},\sigma}(E)= \frac{1}{N} \Sigma_{i,j} e^{\vec{k}
    (\vec{R}_i-\vec{R}_j)}\tilde{S}_{i,j,\sigma}(E)  \\ 
  \tilde{S}_{i,j,\sigma}(E)=\frac{1}{2\pi}\,\int\limits_{-\infty}^{+\infty}dE\,
  e^{-\frac{i}{\hbar}Et} \, \langle [a_{i,\sigma}(t);a_{j,\sigma}^{\dagger}(0)]_+\rangle
\end{multline}
where $[\dots;\dots]_+$ denotes the anticommutator and $\langle\dots\rangle$ the
thermodynamic average. The construction operators are taken to be in the
Heisenberg time-dependent picture.

In an exact spectral-moment analysis in the limit $U\rightarrow\infty$, Harris
and Lange have shown that the spectral density essentially
consists of a two-peak structure \cite{HL67}. The effective Hubbard model (\ref{effham})
must be considered in the strong-coupling regime. Therefore it seems
appropriate to make the following ansatz for the spectral density:
\begin{equation}
  \tilde{S}_{\vec{k},\sigma}(E)=\sum_{j=1,2}\hbar\,\tilde{\alpha}_{\vec{k},\sigma}^{(j)}\,\delta(E-\tilde{E}_{\vec{k},\sigma}^{(j)})
\end{equation}
The still unknown parameters  $\tilde{E}_{j,\sigma}$ and $\tilde{\alpha}_{j,\sigma}$, the
quasiparticle energy and spectral weight, can be calculated by use of the
first four moments of the spectral density:
\begin{equation}
  \begin{split}
    \tilde{M}_{\vec{k},\sigma}^{(n)} =& \int\limits_{-\infty}^{+\infty}dE\,E^n\,
    \tilde{S}_{\vec{k},\sigma}(E) =\\
    &\langle  [ \underbrace{[...[a_{\vec{k},\sigma},\tilde{H}]_-,...,\tilde{H}]_-}_{\text{$n$-fold commutator}} ,
    a_{\vec{k},\sigma}^{\dagger} ]_+\rangle 
  \end{split}
\end{equation}
This procedure is identical to the one performed in \cite{HN97a} for the real
Hubbard problem. An explicit description of the calculation is presented there.

Despite its obvious restrictions, e.g. the complete neglection of
quasiparticle damping, the two-pole approximation together with the
moment method to calculate the free parameters is able to describe
the magnetic properties of the Hubbard model surprisingly
well \cite{HN97a,HN97b}. 

As a result one obtains a selfenergy of the following structure:
\begin{multline}
  \label{selfenergy_SDA}
  \tilde{\Sigma}_{\vec{k},\sigma}^{(\eta)}(E)=U\,\langle n_{i,-\sigma}^{(a)}\rangle\\
  \times \frac{E-\tilde{B}_{-\sigma}-\tilde{F}_{\vec{k},-\sigma}}
  {E-\tilde{B}_{-\sigma}-\tilde{F}_{\vec{k},-\sigma}-U(1-\langle n_{i,-\sigma}^{(a)}\rangle)}
\end{multline}
The decisive terms are $\tilde{B}_{-\sigma}$ and $\tilde{F}_{\vec{k},-\sigma}$ which distinguish
this selfenergy from the Hubbard-I solution \cite{Hub63}. These terms, mainly consisting of
higher correlation functions,  may provoke a
spin-dependent shift and/or deformation of the bands and may therefore be
responsible for the existence of spontaneous magnetism \cite{HN97a,PWN97,PHWNpre98}.

The $\vec{k}$-dependent term $\tilde{F}_{\vec{k},-\sigma}$ seems to be of minor
importance for the magnetic behaviour. Since $\Sigma_{\vec{k}}
\tilde{F}_{\vec{k},-\sigma}=0$, it does not change the center of gravity of the density
of states being mainly responsible for a deformation and narrowing of the bands.
We have neglected this term in the following calculations. A detailed inspection of the influence
of this term, as well as quasiparticle damping will be the subject of future investigations.

The term $\tilde{B}_{-\sigma}$ has the following structure:
\begin{multline}
  \label{bandshift}
  \tilde{B}_{-\sigma}=\Delta_0+\frac{1}{\langle
    n_{i,-\sigma}^{(a)}\rangle(1-\langle
    n_{i,-\sigma}^{(a)}\rangle)}b_{-\sigma}\\ 
  \begin{split}
    b_{-\sigma}=& \frac{1}{N}\sum_{i,j}^{i\neq j}
    \Delta_{i,j} \langle a_{i,-\sigma}^{\dagger}a_{j,-\sigma}(2n_{i,\sigma}^{(a)}-1)\rangle\\
    =&\frac{1}{N}\int\limits_{-\infty}^{+\infty}\sum_{\vec{k}}f_-(E)
    \left(\Delta_{\vec{k}}(\eta)-\Delta_0\right)\\
    &\left(\frac{2}{U} (E-\Delta_{\vec{k}}(\eta))-1\right)
    \tilde{S}_{\vec{k},\sigma}(E-\mu)
  \end{split}
\end{multline} 
$\Delta_0$ denotes the center of gravity of the effective electron hopping
$\Delta_{i,j}$ (see equation (\ref{effmed1})). In the case of a $\vec{k}$-independent, real
selfenergy, as we have obtained with our approximation, the spectral
density can be expressed as:
\begin{equation}
  \label{spec_dens2}
\tilde{S}_{\vec{k},\sigma}(E)=\hbar
\delta\left(E-\Delta_{\vec{k}}(\eta)+\mu-\tilde{\Sigma}_{\sigma}^{(\eta)}(E)\right) 
\end{equation}
Together with the spectral theorem:
\begin{equation}
  \langle n_{\sigma}^{(a)}\rangle =
  \frac{1}{N}\int\limits_{-\infty}^{+\infty}\sum_{\vec{k}}f_-(E) \tilde{S}_{\vec{k},\sigma}(E-\mu)
\end{equation}
the equations (\ref{bandshift}), (\ref{selfenergy_SDA}) and (\ref{spec_dens2})
form a closed set of equations which has to be solved selfconsistently.

After retransforming from the effective to the real medium according to
equations (\ref{equivgreen}) and (\ref{equivself}) respectively, one finally
obtains a solution for the periodic Anderson model.

\section{Results}
In the following sections, the results of the above developed theory
will be presented. The system is determined by the following parameters:
The free conduction band is described by a semi-elliptic density of
states of the width $W=1$ and center of gravity at $E=0$, thus defining
the energy scale.  The $f$-level is characterized by the parameters
$e_f$, its relative position with respect to the center of mass of the
conduction band, and the on-site Coulomb interaction $U$. The
hybridization $V$ is taken as a relatively small parameter, in all cases
examined below, we chose $V=0.2$. Of crucial importance is the total
number of electrons per lattice site, $n_{tot}=n_f+n_s$. Calculations
will be made for the ground state ($T=0$) and for finite temperatures.
\subsection{The paramagnetic system}
\begin{figure}[h]
  \begin{center}
    \epsfig{file=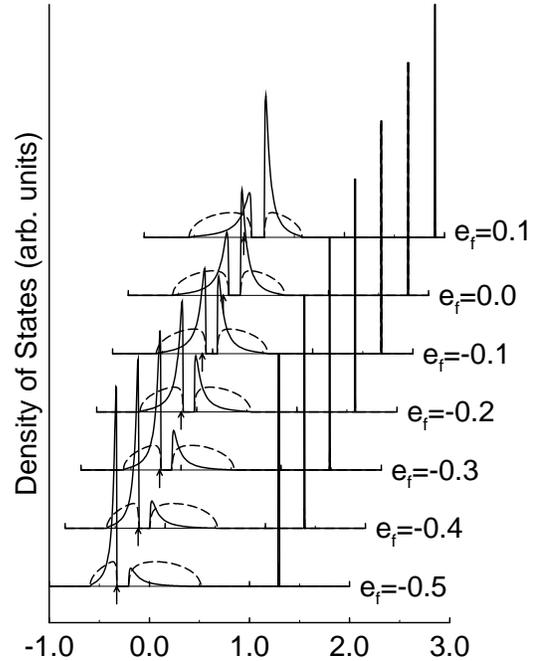,width=8cm}
    \caption{$s$-(dashed line) and $f$(solid line)-density of states. The parameters are $U=1.5$,
      $V=0.2$, $T=0$ and $n_{tot}=1.5$. The chemical potential $\mu$ is
      indicated by the arrows.}
    \label{fig:para_qdos}
  \end{center}
\end{figure}

In figure \ref{fig:para_qdos} the paramagnetic quasiparticle density of
states is plotted for different values of $e_f$. The interaction
constant is chosen to be $U=1.5$, the total occupation number is
$n_{tot}=1.5$. The arrows point to the position of $\mu$, the chemical
potential.

It can be clearly seen that the density of states is divided into three
structures. With the $f$-level positioned within the free conduction band, the
lower two structures exhibit band-like behavior, while the third structure,
the upper ``Hubbard-band'' of the $f$-level, retains its atomic shape. 

The gap which separates the lower two peaks, is induced by the
hybridization. Its position is related to the position of the
free $f$-level, and its
size scales with the hybridization strength $V$.

Contrary to the upper ``Hubbard-band'' which is almost of pure
$f$-character, these lower structures show strong mixing of $s$- and
$f$-states. For $e_f$ approaching one of the band-edges of the free $s$-band, a
strong peak of $f$-states develops. 

In the example shown in figure \ref{fig:para_qdos}, there is always a
peak close to the Fermi energy. This coincidence originates from the
chosen system parameters, more precisely from the number of electrons
$n_{tot}$. \\ Due to the simple concept of the two-pole approximation
used for solving the effective Hubbard model, we do not see a Kondo
resonance. A Kondo resonance would strongly influence the low energy
properties of the system. But as already discussed in the theory
section, for describing the magnetic properties, the correct high energy
behaviour of the selfenergy is of importance
\cite{PHWNpre98}. Therefore, we do believe in our results concerning
magnetism at least for situations and temperatures where a Kondo effect
is not important.
\subsection{The magnetic ground state properties}
\begin{figure}[h]
  \begin{center}
    \epsfig{file=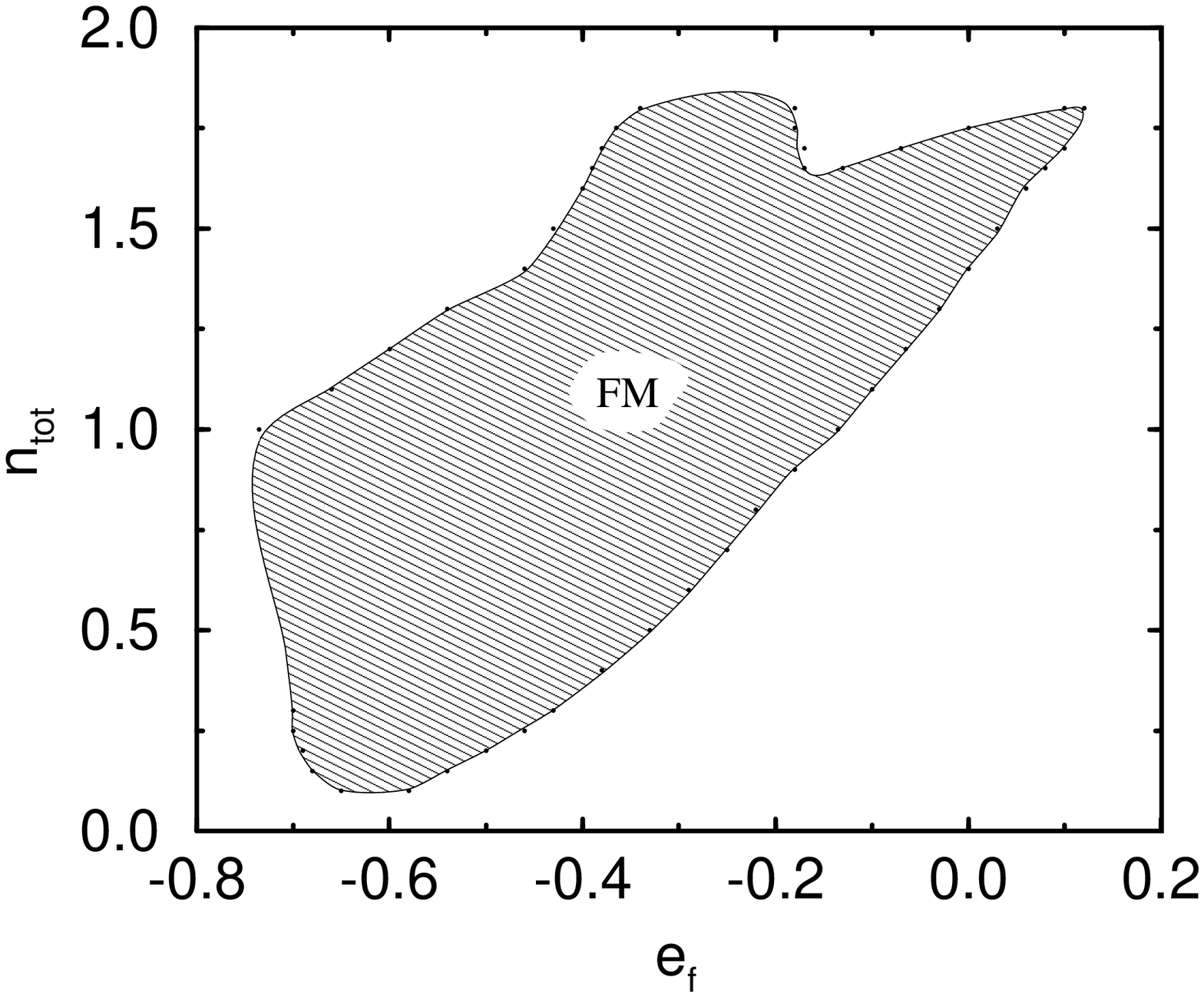,width=8cm}
    \caption{Magnetic phase diagram for the periodic Anderson model at
      $T=0$ with $U=10$ and $V=0.2$. The shaded region is ferromagnetic, the
      surrounding paramagnetic.}
    \label{fig:phdia}
  \end{center}
\end{figure}

In figure \ref{fig:phdia}, we have drawn a magnetic phase diagram of the
periodic Anderson model derived within the above developed theory. This
phase diagram consists of an area where ferromagnetism is stable whereas
in the remaining parameter space ($n_{tot}$ and $e_f$ being the
variables) the ground state is paramagnetic. Antiferromagnetic ordering
was not considered.

For a fixed total number of electrons $n_{tot}$ ferromagnetism can only
exist if the $f$-level lies in a certain range somewhere around the
lower half of the conduction band. For energetically lower positions of
the $f$-level, the ferromagnetic solution of the set of equations still
exists. But since the paramagnetic solution has a lower energy, the
ground state is paramagnetic.  With the $f$-level approaching higher
energies relative to the conduction band, the number of electrons in
$f$-states, $\langle n_f\rangle$ reduces, and the $f$-density of states
broadens. It is commonly accepted that both effects reduce the stability
of ferromagnetism in the system. So it is not surprising to see the
system become paramagnetic with higher values of $e_f$.

\begin{figure}[htbp]
  \begin{center}
    \epsfig{file=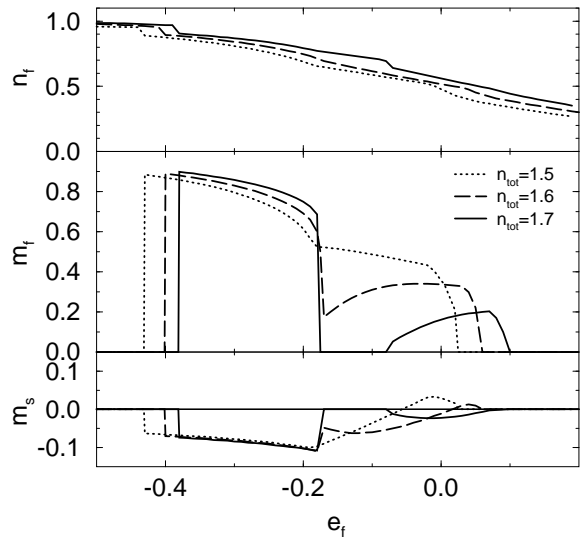,width=8cm}
    \caption{$f$- and $s$-magnetization and the occupation number $n_f$ of the
      $f$-level as functions of $e_f$ for different total occupation numbers
      $n_{tot}$. ($U=10$, $V=0.2$, $T=0$)}
    \label{fig:m_ef}
  \end{center}
\end{figure}
In the following, we want to focus on the intermediate valence regime
(IV) of the periodic Anderson model. Intermediate valence is
characterized by non-integer values of $n_f$, the $f$-level occupation
number. Per definitionem, an ``atomic-like'' level can contain
either none, one or two electrons. Due to the rather high value of $U$,
the local part of the Coulomb repulsion, double occupancy is
suppressed. So a non-integer value of $n_{tot}$ describes mainly fluctuations
between the non-occupied and the single-occupied state. The condition
for IV is found for $e_f$ lying within the conduction band, more
specific, the closer $e_f$ and $\mu$ are, the higher are the charge
fluctuations within the $f$-states, the average $n_f$ occupation number
is lower.

For describing rare earth materials, a total occupation number $n_{tot}$
lower than $n_{tot}=1$ is not of interest. So we will confine our
investigations to the parameter range which builds up the upper right
hand side of the phase diagram in figure \ref{fig:phdia}. In this region, there is 
a ``bay'' in the ferromagnetic isle. For a total number of electrons
around $1.75$, ferromagnetism first disappears with rising $f$-level, but
reappears for little higher $f$-levels, until the system finally becomes
paramagnetic again.

This is shown in more detail in figure \ref{fig:m_ef}, presenting the
magnetization as a function of $e_f$ for three different electron
densities $n_{tot}=\{1.5, 1.6, 1.7 \}$. For $n_{tot}=1.7$ the two
separate regions of stable ferromagnetism can be seen, but already for
the lower occupation numbers, two distinct regions of ferromagnetism can
be distinguished from each other.

The first of these, appearing for low values of $e_f$, sets in with a
first order phase transition at $e_{f,\text{crit}}^{\text{lower,1}}$ and
is stable up to $e_{f,\text{crit}}^{\text{upper,1}}$. The existence of a
first order phase transition seems unusual. But since
$e_{f,\text{crit}}^{\text{lower,1}}$ is a quantum critical point, not a
thermodynamic critical point, first order phase transitions are not
forbidden. At least for  transitions from antiferromagnetic to paramagnetic states, first
order phase transitions were experimentally observed \cite{Sea96}.

This region, usually called Kondo regime, is characterized by almost half filled
$f$-levels forming relatively large local moments. According to the
Schrieffer-Wolff-transformation \cite{SW66}, one expects to find
antiferromagnetic coupling between the $f$-moments and the conduction
band moments. This
behaviour is clearly seen in figure \ref{fig:m_ef}. But the uncorrelated
conduction band shows only a weak magnetic polarization. Thus, the indirect interaction between the local
moments via the conduction band is responsible for the magnetic
ordering. This region of ferromagnetism could be called ``local moment
magnetism'' (LMM), since the electrons forming the relativly large local 
moments are energetically far away from the Fermi energy. Thus, they do
not participate in transport phenomena.

\begin{figure}[htbp]
  \begin{center}
    \epsfig{file=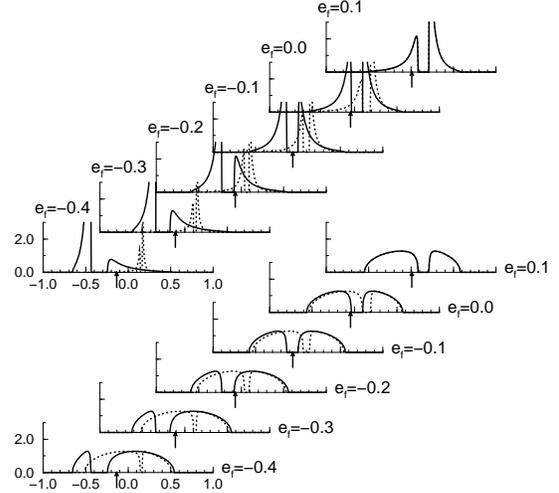,width=8cm}
    \caption{$f$- ($s$-) quasiparticle density of states for different values of
      $e_f$ in the upper (lower) row (full line:
      spin-$\uparrow$, dotted line: spin-$\downarrow$). The arrow denotes the position of the chemical
      potential. The total number of electrons is $n_{tot}=1.5$, the other
      system parameters are $V=0.2$, $T=0$, and $U=10$.}
    \label{fig:qdos_n1.5}
  \end{center}
\end{figure}

The second region of ferromagnetic ordering belongs to the intermediate
valence regime of the PAM. This is specified by $f$-level occupation
numbers $n_f$ away from half filling, and a typical broadening of the
$f$-levels to bands. The average magnetic moment at each $f$-site is
strongly reduced. Furthermore, the coupling between the $f$-levels and the
conduction band is not always antiferromagnetic, but, depending on the
system parameter, can be ferromagnetic. So, intermediate valence
magnetism (IVM) is in many aspects different from the LMM. 

The crossover from the LMM into the IVM regime depends on the
electron density. As already mentioned, for higher $n_{tot}$,
ferromagnetic order disappears at $e_{f,\text{crit}}^{\text{upper,1}}$
to reappear at $e_{f,\text{crit}}^{\text{lower,2}}$ in a second order
phase transition. For $n_{tot}=1.6$, the two critical points coincide,
i.e. $e_{f,\text{crit}}^{\text{upper,1}}=e_{f,\text{crit}}^{\text{lower,2}}$,
thus an immediate transition between the two regions of ferromagnetism is
seen. For even lower occupation numbers, e.g. $n_{tot}=1.5$ there is a
crossover region with linear behaviour of the magnetization.
 
\begin{figure}[htbp]
  \begin{center}
    \epsfig{file=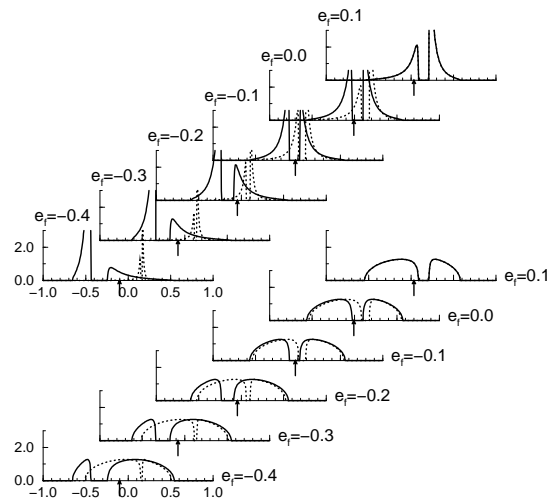,width=8cm}
    \caption{Same as figure \ref{fig:qdos_n1.5}, but $n_{tot}=1.6$.}
    \label{fig:qdos_n1.6}
  \end{center}
\end{figure}

\begin{figure}[htbp]
  \begin{center}
    \epsfig{file=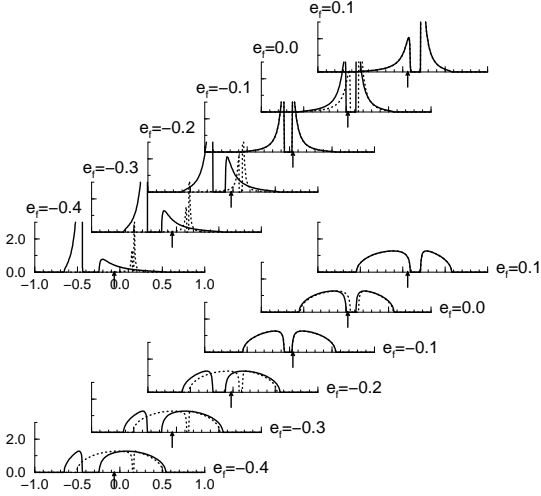,width=8cm}
    \caption{Same as figure \ref{fig:qdos_n1.5}, but $n_{tot}=1.7$.}
    \label{fig:qdos_n1.7}
  \end{center}
\end{figure}
The reason for these different kinds of crossover behaviour can be found
in the quasi particle densities of state (QDOS), as shown in figures
\ref{fig:qdos_n1.5} to \ref{fig:qdos_n1.7}. These QDOS correspond to the
three cases presented in figure \ref{fig:m_ef} with
$n_{tot}=\{1.5,1.6,1.7\}$, respectively. In all three cases, the
crossover region coincides with the region where the chemical potential
$\mu$ crosses the hybridization gap. In case of ferromagnetism, the
hybridization gap of the majority spin direction band is decisive.
 
For $n_{tot}=1.5$ (figure \ref{fig:qdos_n1.5}), the chemical
potential lies in the flat region of the minority band, thus a small
change of $e_f$ will, for constant $n_{tot}$, have only minor effects on 
the quasiparticle densities of states.\\
At $n_{tot}=1.6$ (figure \ref{fig:qdos_n1.6}), the different behaviour is provoked by the
fact that the spin-$\downarrow$ band has a rather sharp peak around
$\mu$, so already small parameter changes can have a strong impact on the system.\\
Finally, for $n_{tot}=1.7$ (figure \ref{fig:qdos_n1.7}), ferromagnetism
breaks down.

It is worth to mention the possibility of a
metal-insulator transition due to the hybridization gap. For
$n_{tot}=1.7$, we see such a transition for $e_f\approx0.1$ as the
system becomes paramagnetic. On the contrary, for $n_{tot}=1.5$ or
$n_{tot}=1.6$, the system stays ferromagnetic during the transition, and 
only spin-$\downarrow$-electrons can contribute to conductivity. But no
metal-insulator transition is seen in this case.

Another fairly striking result of our calculations is the possibility of 
ferromagnetic coupling between the conduction band and the
$f$-states. 

The Schrieffer-Wolff transformation maps the PAM onto the
Kondo model \cite{SW66}. This model considers only a spin-exchange interaction
between the conduction band and the system of $f$-levels. The effective
coupling between the different electron systems 
is antiferromagnetic. So, one would expect to find the conduction band
spin-polarized anti-parallel to the $f$-states. 

\begin{figure}[htbp]
  \begin{center}
    \epsfig{file=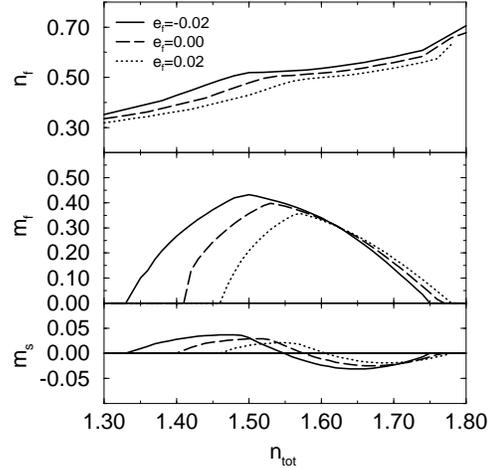,width=7cm}
    \caption{$s$- ($f$-) magnetization $m_s$ ($m_f$) and the $f$-occupation
      $n_f$ as a function of the total occupation number $n_{tot}$. In all
      four cases the free $f$-level $e_f$ is close to the center of mass of
      the free conduction band. The other system parameters are $U=10$,
      $V=0.2$ and $T=0$.}
    \label{fig:m_n}
  \end{center}
\end{figure}

As already mentioned above, we do find parameter sets, where the
conduction band and the localized levels are ferromagnetically
coupled. This can be seen in figure \ref{fig:m_ef}, and in more detail
in figure \ref{fig:m_n}. There, the $f$- and $s$-magnetization as well
as the $f$-occupation number $n_f$ is plotted as a function of
$n_{tot}$ for three different values of
$e_f$ all very close to the center of gravity of the free conduction
band, thus belonging to the intermediate valence
regime. Dependent on $n_{tot}$ the conduction band is polarized either
in the same spin direction as the $f$-system (ferromagnetic coupling),
or in opposite direction (antiferromagnetic coupling).
For higher $n_f$, the
tendency is clearly towards antiferromagnetic coupling. 

If one keeps in mind that the Schrieffer-Wolff transformation is only
valid in the Kondo limit of the PAM, i.e. for half filled $f$-levels,
no discrepancy is found.

\begin{figure}[htbp]
  \begin{center}
    \epsfig{file=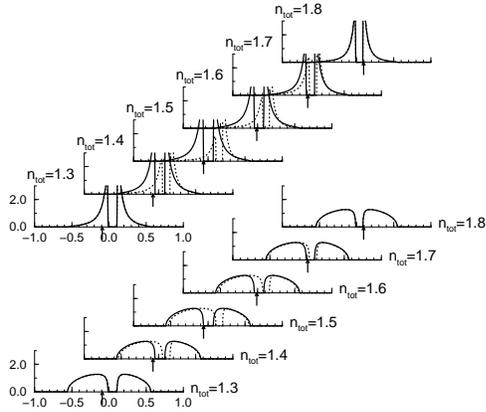,width=7cm}
    \caption{$f$- ($s$-) quasiparticle density of states for different values of
      $n_{tot}$ in the upper (lower) row (full line:
      spin-$\uparrow$, dotted line: spin-$\downarrow$). The arrow denotes the position of $\mu$, the chemical
      potential. The free $f$-level is located at $e_f=-0.02$, the other
      system parameters are $V=0.2$, $T=0$, and $U=10$.}
    \label{fig:qdos_ef-0.02}
  \end{center}
\end{figure}

\begin{figure}[htbp]
  \begin{center}
    \epsfig{file=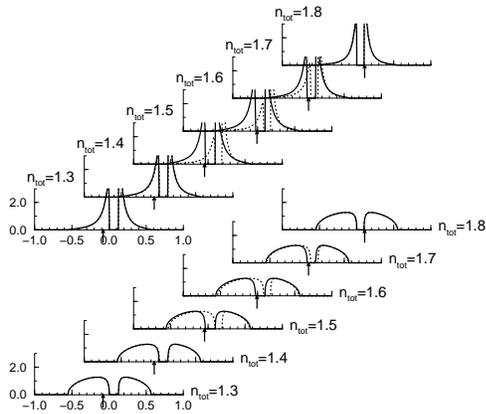,width=7cm}
    \caption{Same as figure \ref{fig:qdos_ef-0.02}, but with $e_f=0.00$.}
    \label{fig:qdos_ef0.00}
  \end{center}
\end{figure}

\begin{figure}[htbp]
  \begin{center}
    \epsfig{file=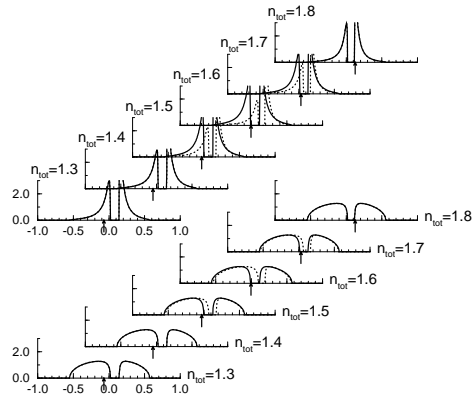,width=7cm}
    \caption{Same as figure \ref{fig:qdos_ef-0.02}, but with $e_f=0.02$.}
    \label{fig:qdos_ef0.02}
  \end{center}
\end{figure}

The mechanism which controls the direction of spin-polarization of the
conduction band can be seen in the densities of state (figures
\ref{fig:qdos_ef-0.02} to \ref{fig:qdos_ef0.02}). While for
$\rho_{f,\sigma}(E)$ the weight in each of the quasiparticle bands is
distributed very asymmetric, $\rho_{s,\sigma}(E)$ is always rather
flat. But due to the hybridization, the partial $s$-electron QDOS
expierences the same spin-dependent bandshift as the respective
$f$-QDOS. So, especially, when the lower band is not yet filled
(i.e. low $n_{tot}$), the spin-$\uparrow$ $s$-band outweighs the
respective spin-$\downarrow$ band. Beginning with $\mu$ crossing the
upper band edge of the lower band, the spin-$\downarrow$ $s$-band gets
filled, thus the QDOS induced tendency towards ferromagnetic coupling
gets reduced and the antiferromagnetic exchange dominates.

The antiferromagnetic exchange expresses itself in $\rho_{s,\sigma}(E)$ by the fact that for spin $\uparrow$ the lower of the hybridization bands has less, 
for spin $\downarrow$ more weight than the upper hybridization band.

The change of sign of the conduction band polarization provokes another
interesting question. There is a dedicated set of parameters for which
the $s$-band is not polarized at all, but the $f$-moments order
ferromagnetically. This suggests that the spin polarization of the
$s$-system is not responsible for the magnetic ordering of the
$f$-levels. 

In the LMM regime, the situation is clear. A strong antiferromagnetic spin exchange
between localized levels and conduction band states leads to an indirect 
coupling of the $f$-moments, thus to ferromagntic ordering.

On the contrary, in the IVM regime, ferromagnetism in the $f$-system is
obviously independent of a magnetic ordering in the conduction
band. Otherwise one would expect to see some effects on the
$f$-magnetization at the critical occupation number $n_{tot}$ where
$m_s=0$. In figure \ref{fig:m_n} nothing like this can be observed. So
in our opinion, spin exchange processes between the conduction band and
the localized levels cannot be the major effect for producing the
ferromagnetic ordering in the $f$-system. Of course, spin fluctuations
are still present, thus inducing the magnetic ordering in the conduction
band.  But there must be a different mechanism leading to the
ferromagnetic ordering of the $f$-moments. We propose a single band
effect within an effective $f$-band. Due to charge fluctuations between
$f$- and $s$-system, an effective $f$-$f$-hopping is possible. This
leads to the formation of a narrow band analogously to a single-band
Hubbard model. In the latter, ferromagnetism is, at least under certain
conditions, expected to exist \cite{Ulm98,Taspre,Weapre97,Veapre97},
favored by a rather small band which should show a divergence at a band
edge, and an occupation number not too far away from half filling. By
examining the $f$-electron QDOS (see for example figure
\ref{fig:qdos_ef-0.02}, one sees that these conditions are fullfilled in
the effective $f$-band.
\subsection{Finite Temperatures}
When examining magnetism, the temperature dependence of several
quantities is of major interest. In this section, the magnetization
curves as well as the Curie-temperatures $T_c$ for different system
parameters are presented.

\begin{figure}[htbp]
  \begin{center}
    \epsfig{file=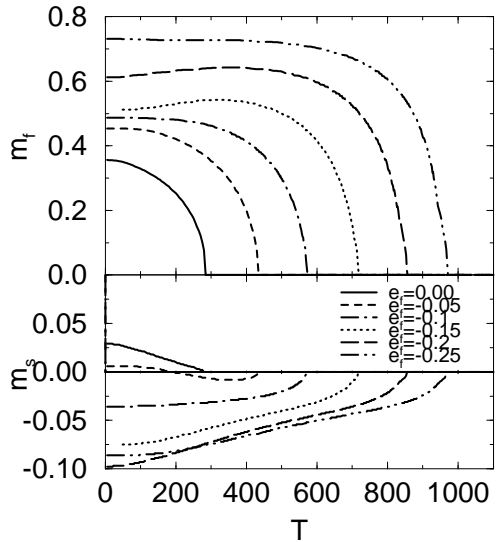,width=7cm}
    \caption{$s$- and $f$-magnetization as function of the temperature
      $T$ for $n_{tot}=1.5$, $U=10$ and $V=0.2$. The temperature scale is
      explained in the text.}
    \label{fig:m_t_n1.5}
  \end{center}
\end{figure}

\begin{figure}[htbp]
  \begin{center}
    \epsfig{file=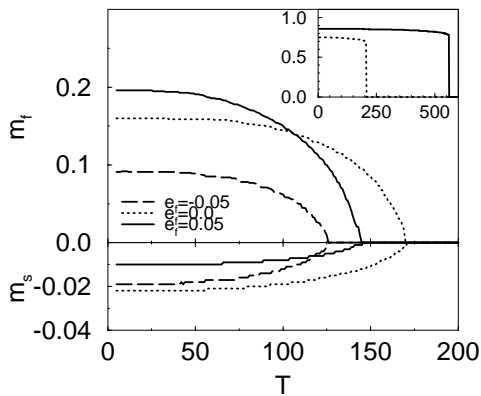,width=7cm}
    \caption{Same as figure \ref{fig:m_t_n1.5}, but for $n_{tot}=1.7$. In
      the inset, the $f$-magnetization is shown for $e_f=-0.3$ (solid
      line) and $e_f=-0.2$ (dotted line).}
    \label{fig:m_t_n1.7}
  \end{center}
\end{figure}

When the bandwidth of the free conduction band is taken to be $W=1$, the
temperatures are given in Kelvin. The figures \ref{fig:m_t_n1.5} and
\ref{fig:m_t_n1.7} show the temperature dependence of the $s$- and
$f$-magnetization for two electron densities and various $f$-level
positions $e_f$. The $T=0$-properties for these parameters were
presented in figure \ref{fig:m_ef}.

For $n_{tot}=1.5$, as presented in figure \ref{fig:m_t_n1.5}, the
magnetization vanishes continuously at $T_c$, whereas for $n_{tot}=1.7$
(see fig. \ref{fig:m_t_n1.7}) two different behaviours can be seen
according to the two different regions of magnetism: the local moment
magnetism (LMM) and the intermediate valence magnetism (IVM).  For low
$e_f$, i.e. in the LMM regime (see inset of figure \ref{fig:m_t_n1.7}),
the phase transition is of first order, for $e_f$ in the IVM regime of
second order. So the two distinct regions of ferromagnetism seen in
figure \ref{fig:m_ef} show a different temperature dependent behaviour.

In figure \ref{fig:m_t_n1.5}, two interesting features can be
observed. There can be a temperature driven increase in
$f$-magnetization, and the magnetic polarization of the
conduction band can change sign as function of temperature.

A magnetization increasing with temperature seems to be quite
extraordinary, and at first sight, unplausible for thermodynamic
reasons. But on second sight, even a temperature-driven spontaneous
ordering from a paramagnetic to a ferromagnetic state is
possible \cite{VIIK88}. In our case, the system does not undergo a phase
transition, only the magnetization increases slightly for certain
parameters. This can be easily understood as a quasiparticle densities
of states effect: If the system parameters are chosen, such that the
lower edge of the upper hybridization spin $\uparrow$-band is close to,
but still above the Fermi energy, these states are unoccupied for
$T=0$. But already a small softening of the Fermi function can induce a
strong increase in occupied spin-$\uparrow$ states since this band edge
generically consists of a rather sharp peak. The ratio of
spin-$\uparrow$ electrons to spin-$\downarrow$ electrons increases, the
magnetization rises.

The other feature is observed in the lower picture of figure
\ref{fig:m_t_n1.5} showing the conduction band polarization. As already
seen for $T=0$, it can either be parallel (ferromagnetic coupling) or
antiparallel (antiferromagnetic coupling) to the $f$-magnetization, but
sometimes, e.g. for $e_f=-0.05$, it can also cross the $m_s=0$-axis at
some temperature $T^{\star}$. Quite remarkable is the behaviour of the
$f$-magnetization which is continuous even at $T=T^{\star}$. This
resembles the $T=0$ behaviour of the magnetization as function of
$n_{tot}$ in the IV regime discussed in the previous section.  Again, it is
indicated that the coupling mechanism responsible for the alignment of
the $f$-moments in the intermediate valence regime does not depend on
the explicit polarization of the conduction band. This polarization has
to be understood as a consequence, and not the cause of the ordering of
the $f$-system.

\begin{figure}[htbp]
  \begin{center}
    \epsfig{file=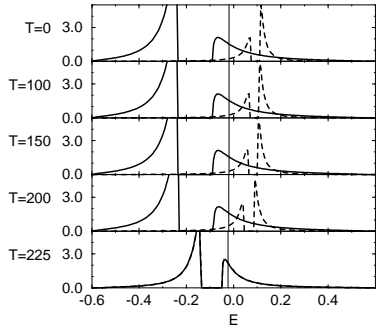,width=7cm}
    \caption{$f$-electron quasiparticle density of state (QDOS) as a
      function of energy for different temperatures (solid line: spin
      $\uparrow$, dashed line: spin $\downarrow$). The system
      parameters are $n_{tot}=1.7$, $e_f=-0.2$, $U=10$ and $V=0.2$. The
      corresponding magnetization curve is presented in the inset of
      figure \ref{fig:m_t_n1.7}. The system shows a first order phase
      transition at $T_c\approx 208$.}
    \label{fig:qdos_t_ef-0.2}
  \end{center}
\end{figure}
\begin{figure}[htbp]
  \begin{center}
    \epsfig{file=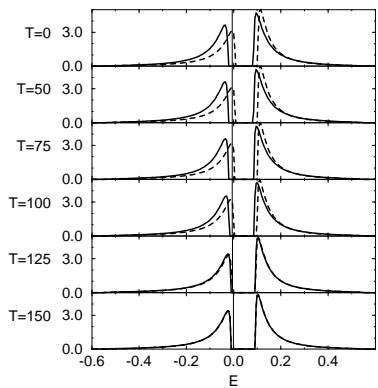,width=7cm}
    \caption{same as figure \ref{fig:qdos_t_ef-0.2}, but for
      $e_f=-0.05$. This corresponds to the dashed line in figure
      \ref{fig:m_t_n1.7}, with $T_c\approx 114$.}
    \label{fig:qdos_t_ef-0.05}
  \end{center}
\end{figure}

In the figures \ref{fig:qdos_t_ef-0.2} and \ref{fig:qdos_t_ef-0.05}, the 
demagnetization behaviour is examined by means of the quasi particle
densities of states. These illustrate both the first order transition in
the LMM regime ($e_f=-0.2$: figure \ref{fig:qdos_t_ef-0.2}) and the second order
transition in the IV region ($e_f=-0.05$: figure \ref{fig:qdos_t_ef-0.05}).
In both cases, the demagnetization is characterized
by two effects, a bandshift and a weight transfer mechanism. The
bandshift between spin $\uparrow$ and spin $\downarrow$ band decreases
with increasing temperature while at the same time a transfer of
spectral weight between the hybridization bands occurs. For the spin
$\uparrow$-bands, weight is transferred from the lower to the upper
band, for the spin $\downarrow$-bands just the contrary takes place.                                

For $e_f=-0.2$, ferromagnetism breaks down at $T=T_c\approx 208$, we see 
a first order phase transition. Contrarily, for $e_f=-0.05$, the
demagntization behaves smoothly up to $T=T_c\approx
114$, thus a second order phase transition occurs.

\begin{figure}[htbp]
  \begin{center}
    \epsfig{file=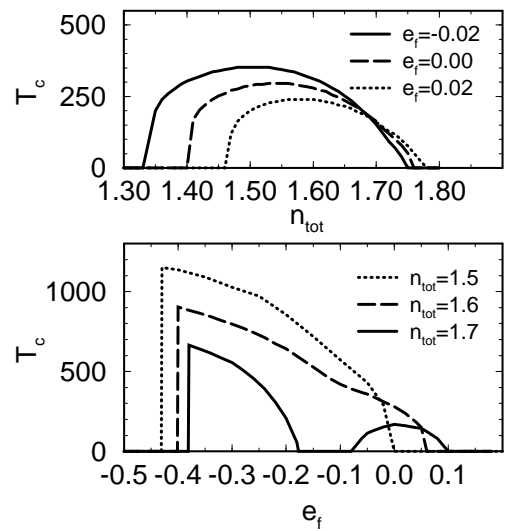,width=7cm}
    \caption{Curie-temperature $T_c$ in the upper picture, as function
      of $n_{tot}$, and in the lower as function of $e_f$ ($U=10$ and $V=0.2$)}
    \label{fig:tc}
  \end{center}
\end{figure}

Finally, in figure \ref{fig:tc}, the Curie-temperatures as function of
$n_{tot}$ (upper figure) and $e_f$ (lower figure) are presented.  The
continuous vanishing of $T_c$ as a function of $n_{tot}$ corresponds to
the second order phase transitions seen in figure \ref{fig:m_n}.

Comparing figure \ref{fig:m_n} and the Curie temperature as a function
of $n_{tot}$ in the IV regime (figure \ref{fig:tc}), one observes that
the maximum of $T_c(n_{tot})$ coincides neither with the maximum of
$m_f(n_{tot})$ not the maximum of $m_s$ at $T=0$. Taking the
Curie temperature as a measure of the thermodynamic stability of
ferromagnetism, this indicates again that the magnetic polarization of
the conduction band does not directly cause the magnetic ordering of the
$f$-moments.  Thus, we conclude that the magnetic polarization of the
conduction band is rather induced by the magnetically ordered atomic
levels than responsible for the ordering itself. This supports our
proposition that the magnetic ordering of the $f$-moments in the
intermediate valence regime is rather a single band effect within the
$f$-system than caused by a spin-exchange between $f$-levels and
conduction band.

In figure \ref{fig:tc} showing $T_c$ as a function of $e_f$, the splitting of the two
regions of ferromagnetism for $n_{tot}\gtrapprox 1.7$ can be seen
again. The lowest boundary of ferromagnetism with respect to $e_f$ is
given by the first order phase transition at the quantum critical point
$e_{f,\text{crit}}^{\text{lower},1}$. Right above this critical point,
both the Curie temperature as well as $m_f(T=0)$ show their
maximum. Here, in the LMM regime, the Schrieffer-Wolff transformation
suggests that an effective coupling of $f$-moments is possible only via 
an (antiferro-) magnetic exchange with the conduction band. But again,
for the IVM region, no connection can be made between the $T=0$ moments of the two
electronic subsystems and the Curie temperature $T_c$.

\section{Conclusions}
In this paper, we have presented a new approach to the periodic Anderson
model. Our approximation is based on an exact mapping of the PAM onto an
effective Hubbard model. Since this effective model is clearly located
in the strong coupling regime, a for this limit rather well motivated
approximation,the SDA is used to find self-consistent solutions. Even
though the particular low-temperature properties of the Anderson model
like a formation of a Kondo resonance are not accessible within our
approximation, it is trustworthy concerning magnetism. A major advantage
of this approach is its numerical simplicity, enabling calculations for
the full parameter range and the accessibility of all major quantities
of interest via the selfenergy given on the real axis.

Using this method we have shown the $T=0$ phase diagram and have found 
two distinct regions of ferromagnetism. The first of these is best
described by the term ``local moment magnetism'' (LMM) while the other
located within the intermediate valence region (IVM)
displays itinerant character.

The local moment region is close to the Kondo regime of the PAM. The
magnetic moments within the $f$-system are fully polarized, the coupling 
between $f$-moments and conduction band is antiferromagnetic, as
expected from the Schrieffer-Wolff transformation. The Curie
temperatures are rather high.

On the other side, the magnetism in the intermediate valence regime is
quite weak, in a sense that the average magnetic moment per site is low
and the Curie temperatures are smaller than in the LMM region. The
magnetic polarization of the conduction band can either be parallel or
antiparallel to the $f$-moments, at some isolated points, the conduction 
band is not polarized at all. This means, the polarization of the
conduction band should be of minor importance to the magnetic ordering
of the $f$-moments. 

We conclude that the mechanism resposible for the ferromagnetic ordering 
is different for the LMM and the IVM regime. Whereas in the LMM regime,
a spin-exchange according to the Schrieffer-Wolff transformation causes
ferromagnetic ordering of the $f$-moments, in the IV regime, the spin
fluctuations still lead to a spin polarization of the conduction band,
but rather more important for the ferromagnetic ordering in the
$f$-levels are the
charge fluctuations leading to an effective broadening of the $f$-levels 
to a band. This band has those properties expected to form
ferromagnetism within a single-band Hubbard model. 

\begin{appendix}

\section*{\normalsize \bf Acknowledgement}
One of us (D.M.) gratefully acknowledges the support of the Friedrich-Naumann foundation.

\end{appendix}
\end{document}